\documentclass[twocolumn,showpacs,preprintnumbers,amsmath,amssymb]{revtex4}

\usepackage{graphicx}
\usepackage{dcolumn}
\usepackage{bm}
\usepackage{subfigure}

\begin{document}


\title{Geometry-induced pulse instability in microdesigned catalysts:\\ the effect of boundary curvature}

\author{L. Qiao}
\author{I. G. Kevrekidis}
\altaffiliation[Also at ]{the Program in Applied and Computational
Mathematics(PACM) , Princeton University, Princeton,NJ 08544, USA}
\affiliation{Department of Chemical Engineering, Princeton
University, Princeton, NJ 08544 USA}
\author{C. Punckt}
\author{H. H. Rotermund}
\affiliation{Fritz-Haber-Institut der MPG, Faradyweg 4-6, 14195
Berlin, Germany}

\date{\today}

\begin{abstract}
We explore the effect of boundary curvature on the instability of
reactive pulses in the catalytic oxidation of CO on microdesigned
Pt catalysts.
Using ring-shaped domains of various radii, we find that the
pulses disappear (decollate from the inert boundary) at a turning
point bifurcation, and trace this boundary in both physical and
geometrical parameter space.
These computations corroborate experimental observations of pulse
decollation.
\end{abstract}

\maketitle

\section{Introduction}
\label{intro}


Pattern formation in spatially uniform media has been studied
extensively (see e.g. the review by Cross and Hohenberg
\cite{Cross} and, specifically for catalytic surfaces, the recent
review by Imbihl \cite{Imbihl}).
%
Since the pioneering Photoemission Electron Microscopy (PEEM) work
of Ertl and coworkers in 1989 \cite{peem}, high vacuum CO
oxidation on single crystal Pt catalysts has been a paradigm for
such studies.
%
Current research increasingly focuses on the interplay between
spontaneous pattern formation and spatial medium nonuniformity, as
well as with spatiotemporal variations of the medium properties.
Intentionally designing the geometry of the medium at the
microscopic level has been finding an increasing number of
applications across disciplines in recent years: from the guidance
of cell migration through micropatterning \cite{Whitesides} and
the template-based self-assembly of diblock copolymers (e.g
\cite{Register}) to the observation of binary fluids
\cite{Balazs1} and phase separation in confined geometries (e.g.
\cite{Boltau}).
Examples of complex, microdesigned geometries used in the study of
reacting systems include Belousov-Zhabotinsky (BZ) catalyst
microprinting \cite{Showalter} and CO oxidation on
microlithographically shaped domains on Pt catalysts
\cite{boundaries2}. The study of reactive pattern formation in
random heterogeneous media is a topic of active current modelling
work (e.g. \cite{Panfilov,Glass}); natural heterogeneous media
(such as water-in-oil microemulsions in \cite{Epstein}) are also
being studied.
Microfluidics is another field that hinges on the micron-scale
design of geometry to control flow and transport (see e.g. the
review \cite{Stone}); beyond passive geometry design, extensive
developments in {\it active} addressing, for example in control of
droplet breakup \cite{Joanicot} or micromixing \cite{Stroock} have
been realized.
Actively and spatiotemporally altering the activity of chemically
reactive media is exemplified in \cite{Science2001} for thermally
addressed heterogeneous catalytic reactions and in \cite{Sakurai}
for photosensitive BZ media.
The recent work of \cite{Lucchetta} marries microfluidic
technology with the active addressing of pattern forming reactive
media in a developmental biology context, using temperature fields
to perturb pattern formation in drosophila embryos.

The interplay of nonlinear reaction-diffusion dynamics with the
effects of boundaries has been extensively studied both in theory
and in experiments; a remarkable early observation was that
rotating spiral waves of a BZ reagent could be initiated close to
domain boundaries with sharp corners \cite{boundaries1}.
Observations in catalytic reactions include that (a) below a
certain critical domain size, the frequency of rotating waves in a
small circular domain is affected by the domain size
\cite{boundaries3}; and (b) that the interaction of patterns with
inert or active boundaries causes the pinning, transmission and
boundary backup of spirals, and instability of pulses turning
around corners \cite{boundaries4}.
The presence of boundaries can also give rise to new patterns that
have not been observed in homogeneous reaction-diffusion systems
\cite{boundaries2}.

Travelling pulses in excitable media constitute an important
building block of pattern formation.
They have been documented in a variety of nonlinear dynamical
systems, including CO oxidation on Pt(110), arising from a
localized, finite perturbation of a linearly stable steady state
\cite{pulse1,pulse2}.
The instabilities and bifurcations of solitary travelling pulses
in the CO oxidation on Pt(110) have been well studied in the
one-dimensional case for a wide range of parameters
\cite{pulse_1d1,pulse_1d2}.
A recent study explored the effects of anisotropic diffusion on
two-dimensional pulse propagation in the thin ring limit
\cite{thin_ring}.
However, the experimentally observed pulses in the catalytic CO
oxidation on Pt(110) propagate on a truly two-dimensional surface
and interact with boundaries and heterogeneities of finite size; a
more detailed study of the boundary effects on two-dimensional
pulse propagation has been missing.
The use of a simplified two-variable reaction-diffusion model
\cite{two_variable_model} in previous studies of pulses in CO
oxidation also creates some difficulty in comparing computational
results with experimental data.
Derived from the full, three-variable model \cite{full_model}
through approximations and simplifications, the parameters in the
two-variable model do not have simple, explicit physical meaning;
variation of such model parameters is not easy to realize
experimentally.

In this paper, we use the full, three-variable
Krischer-Eiswirth-Ertl (KEE) model for CO oxidation on Pt(110) to
study and analyze the behavior of travelling pulses in fully
two-dimensional ring structures.
In particular, we perform stability and continuation computations
with respect to both {\it physical parameters} (such as the
partial pressure of CO($P_{CO}$)) and {\it geometric parameters}
such as the ring radii.
The curvature of an inert boundary on which a pulse is attached
plays a crucial role in affecting the pulse dynamics; the ring
geometry provides an ideal setting in which to study this
fundamental interaction since the (constant) curvature is
prescribed and can be varied systematically when constructing the
domain.

The paper is organized as follows:
We begin with a brief introduction to the KEE reaction-diffusion
model for catalytic CO oxidation on Pt(110) in Section
\ref{model}.
Section \ref{discussion} shows the effect of boundary curvature on
the stability of two-dimensional travelling pulses in two distinct
cases: for fixed geometry, the dependence of pulse stability on
the partial pressure of CO is first studied ; then, keeping
$P_{CO}$ constant, we study the dependence of pulse stability on
the boundary curvature ---both inner and outer---.
Representative experimental results supporting the calculation are
presented in Section \ref{experiments}.
We briefly summarize our results and conclude in
Section~\ref{conclusion}.


\section{Modeling}
\label{model}

We use the KEE reaction-diffusion model for CO oxidation on
Pt(110) with a surface phase transition \cite{full_model}.
This surface reaction follows a Langmuir-Hinshelwood mechanism:

\[CO+* \leftrightharpoons CO_{ads}\]
\[2*+O_2 \rightarrow 2O_{ads}\]
\[CO_{ads}+O_{ads} \rightarrow 2*+CO_2\uparrow
\]
accompanied by a $1\times2\rightarrow1\times1$ phase transition of
the Pt(110) surface due to CO adsorption.
The equations in this kinetic model are

\begin{equation}
\dot{u}=k_us_up_{CO}\left(1-\left(\frac{u}{u_s}\right)^3\right)-k_1u-k_2uv+D_u\nabla^2u
\label{eqn_u}
\end{equation}

\begin{equation}
\dot{v}=k_vp_{O_2}(ws_{v_1}+(1-w)s_{v_2})\left(1-\frac{u}{u_s}-\frac{v}{v_s}\right)^2-k_2uv
\label{eqn_v}
\end{equation}

\begin{equation}
\dot{w}=k_3(f(u)-w) \label{eqn_w}
\end{equation}
where $u$, $v$ and $w$ denote the surface coverage of CO and O,
and the surface fraction of the $1\times1$ phase respectively.
The adsorption rate constant for CO and O$_2$, $k_u$ and $k_v$
respectively, are set to be constant within the temperature range
considered in this paper. The rate constants $k_1,k_2$ and $k_3$
for the desorption, reaction and surface phase transition are
given by the Arrhenius formula $k_i=(k^0)_iexp(-E_i/RT)$; $T$ is
the temperature of the single crystal.
The function $f(u)$ has been fitted to experimental data to give
the rate of surface phase transition as a function of $u$, the
coverage of CO, as follows:

\[f(u)=\left\{\begin{array}{ccc}0&\mbox{for}&u\leqslant0.2\\
\frac{u^3-1.05u^2+0.3u-0.026}{-0.0135}&\mbox{for}&0.2<u<0.5\\
1&\mbox{for}&u\geqslant0.5\end{array}\right.
\]
The variable $v$ can, in principle, be adiabatically eliminated to
give an activator-inhibitor type PDE system of two equations
\cite{two_variable_model}.

We discretize the three model equations over a two-dimensional
ring-shaped domain using the Finite Element Method as implemented
in the package FEMLAB (\url{http://www.comsol.com}). FEMLAB
provides a dynamic simulator for the problem; we want to find
stable and unstable stationary (in a travelling frame) states,
perform continuation/bifurcation calculations, and compute the
linearized stability of the solutions we find.
We used matrix-free linear algebra algorithms to perform these
computations, with the right-hand-side of our equations provided
on demand by FEMLAB.
Newton-Krylov GMRES (see e.g. \cite{Kelley}) was used for
stationary state computations, coupled with pseudo-arclength
continuation to trace the bifurcation diagrams; the Jacobian of
the discretized equations was exported from FEMLAB into a MATLAB
code for this purpose.
ARPACK was used within MATLAB for stability computations of the
resulting generalized eigenvalue problem.

Due to the symmetry of the domain, we rewrite the system of
differential equations in a frame which rotates at an angular
speed $\omega$. The model equations written in co-moving frame and
polar coordinates are given by:
\begin{equation}
\dot{u}=k_us_up_{CO}\left(1-\left(\frac{u}{u_s}\right)^3\right)-k_1u-k_2uv+D_u\nabla^2u+\omega\partial_{\theta}u
\label{eqn_co_u}
\end{equation}

\begin{equation}
\dot{v}=k_vp_{O_2}(ws_{v_1}+(1-w)s_{v_2})\left(1-\frac{u}{u_s}-\frac{v}{v_s}\right)^2-k_2uv+\omega\partial_{\theta}v
\label{eqn_co_v}
\end{equation}

\begin{equation}
\dot{w}=k_3(f(u)-w)+\omega\partial_{\theta}w\label{eqn_co_w}
\end{equation}
The diffusion of CO is chosen to be isotropic for simplicity.
By appending an extra pinning condition to remove rotational
invariance, we can find the appropriate {\it co-rotating} speed
$\omega$ as well as the steady pulse shape travelling with this
speed, as steady state solutions to Eqn. \ref{eqn_co_u},
\ref{eqn_co_v} and \ref{eqn_co_w}.
It is worth noting that ways to find the ``right" co-rotating
speed, even during transient evolution, have been devised using a
``template based" approach (\cite{RowleyMarsden,RKML,Beyn}).
%

We want to study the influence, on the stability of the travelling
pulses, of the boundary curvature by varying the ring radii.
The ring radii do not appear explicitly in the equations, however;
they only appear in the boundary conditions.
We therefore first carry out a linear transformation of
coordinates, mapping the physical domain (the strip between the
inner radius $r_{i}$ and outer radius $r_{o}$) to a {\it constant}
computational domain (between the fixed values $\rho_i, \rho_o$)

Let $r,\theta$ be the independent variables in the original polar
coordinates. The diffusion term in Eqn. (\ref{eqn_u}) is expressed
as
\[
\nabla^2u=\frac{1}{r}\frac{\partial{u}}{\partial{r}}+
\frac{\partial^2u}{\partial{r^2}}
+\frac{1}{r^2}\frac{\partial^2{u}}{\partial{\theta^2}},~r\in[r_{i},r_{o}],~\theta\in(0,2\pi]\]
Define another set of polar coordinates($\rho, \phi$)
\begin{eqnarray}
\rho &=& (r-r_{i})\times\frac{\rho_{o}-\rho_{i}}{r_{o}-r_{i}}+\rho_{i},~~~~\rho \in [\rho_i,\rho_o] \nonumber \\
\phi &=& \theta,~~~~~~~~~~~~~~~~~~~~~~~~~~~~~~~~\phi\in(0,2\pi]
\nonumber
\end{eqnarray}
and express the term of CO diffusion in the new polar coordinates:
\begin{eqnarray}
\nabla^2u&=&\frac{\sqrt{c_1c_2}}{\rho}\frac{\partial{u}}{\partial{\rho}}+
c_1\frac{\partial^2u}{\partial{\rho^2}}
+\frac{c_2}{\rho^2}\frac{\partial^2{u}}{\partial{\phi^2}} \nonumber\\
&=&\widetilde{\nabla}\cdot\left(\left[\begin{array}{cc}c_1&0\\0&c_2\end{array}\right]\cdot
\widetilde{\nabla}{u}\right)+\left[\frac{\sqrt{c_1c_2}-c_1)}{\rho},0\right]\cdot\widetilde{\nabla}{u} \nonumber\\
\label{transform}
\end{eqnarray}
where
\[
c_1=\left(\frac{\rho_{o}-\rho_{i}}{r_{o}-r_{i}}\right)^2,~c_2=\left(\frac{\rho}{(\rho-\rho_{i})\times\frac{r_{o}-r_{i}}{\rho_{o}-\rho_{i}}+r_{i}}\right)^2
\]
and $\widetilde{\nabla}u$ is the gradient of $u$ in the new polar
coordinates.
Now $r_o$ and $r_i$ appear explicitly in $c_1,c_2$.
By substituting Eqn. (\ref{transform}) in Eqn. (\ref{eqn_u}), we
obtain a new PDE system that can be discretized and rewritten in
the co-moving frame to solve for the steady pulse solutions.

\section{Computational results}
\label{discussion}

\subsection{Influence of the presence of boundary}
\begin{figure}
\centering
\includegraphics[width=0.9\columnwidth]{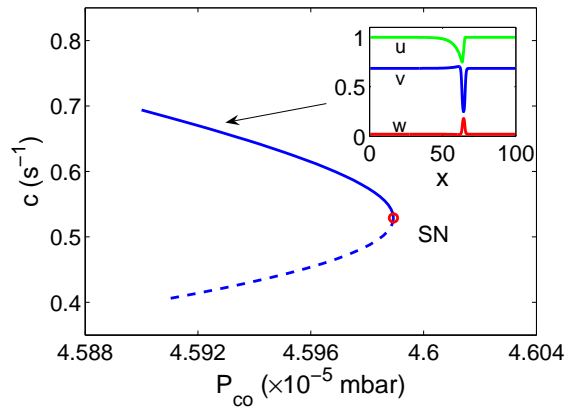}
\caption{Bifurcation diagram of 1D pulses with respect to the
partial pressure of CO. The variable $c$ is the propagating speed
of the steady pulse (a representative one is shown in the inset).
The turning point (saddle-node bifurcation) is indicated by a red
circle. Solid (dashed) lines denote stable (unstable) steady pulse
solutions. $T= 540K, P_{O_2}= 1.33\times10^{-4} mbar$. The inset
shows the typical shape of $u$, $v$ and $w$ at the steady state
marked by the arrow. } \label{1D_bifur}
\end{figure}

\begin{figure}
\centering
\includegraphics[width=0.9\columnwidth]{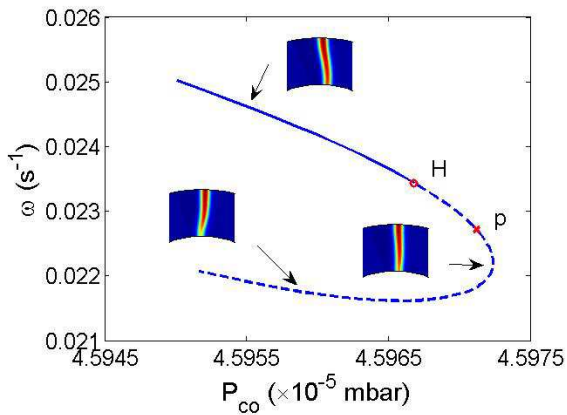}
\caption{Bifurcation diagram of pulses in a 2D ring showing the
influence of the CO partial pressure on the angular speed
($\omega$) of the travelling pulse at stationarity.
Point $H$, marked with a red circle, denotes a Hopf bifurcation.
Solid (dashed) lines denote stable (unstable) steady pulse
solutions.
The insets show profiles of O coverage.
The color is scaled from blue to red according to the coverage of
oxygen from low to high.
The leading eigenvalues and corresponding eigenvectors at points
$H$ and $p$ are plotted in Fig. ~\ref{eigmode_pco}.
The inner and outer dimensionless radius of the ring are 20 and 30
respectively.
$T= 540K, P_{O_2}= 1.33\times10^{-4} mbar; P_{CO}$ is the
bifurcation parameter.} \label{bifur_pco}
\end{figure}

\begin{figure}
\centering
\includegraphics[width=0.9\columnwidth]{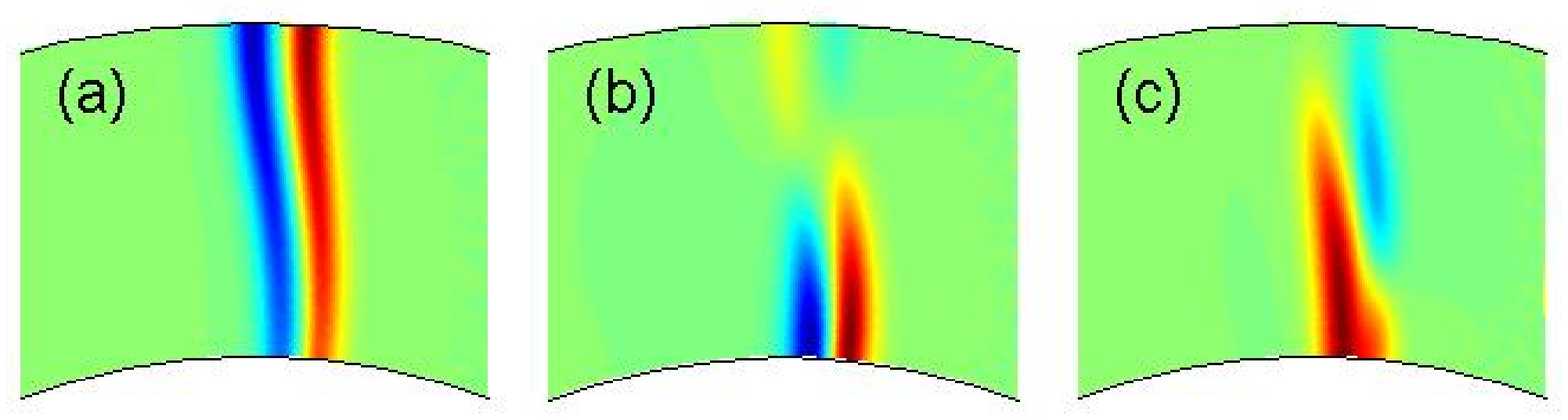}
\includegraphics[width=0.9\columnwidth]{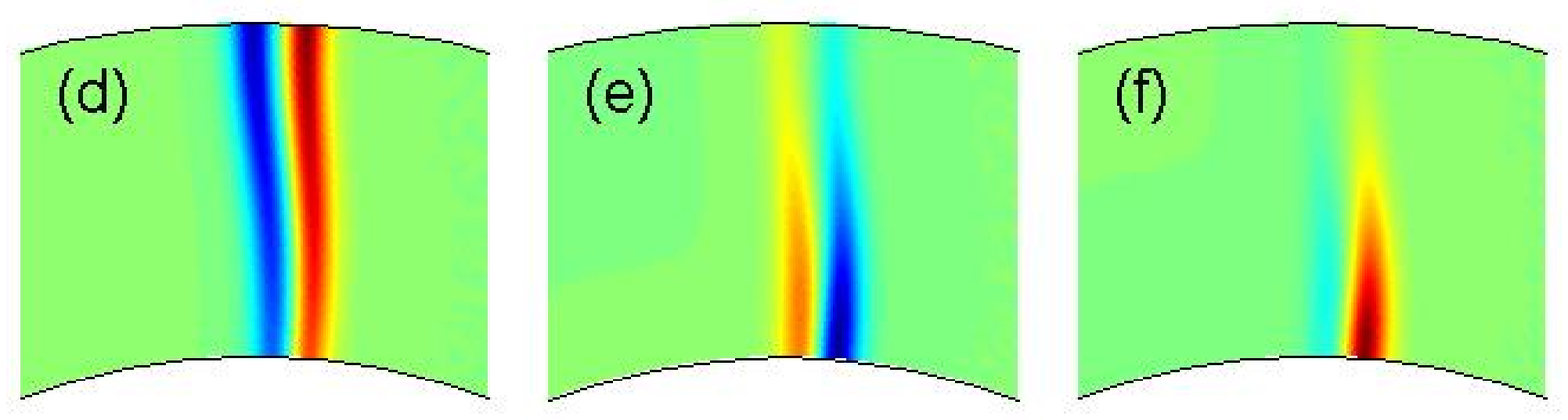}
\caption{The corresponding eigenmodes of three leading eigenvalues
at point $H$ and $p$ as in Fig. ~\ref{bifur_pco}. (a)-(c) are
eigenmodes for point $H$. (a) $\lambda=0.0005$, (b) and (c)
$\lambda=-0.0006\pm0.2591i$. (d)-(f) are eigenmodes for point $p$.
(d) $\lambda=0.0026$, (e) $\lambda=0.1172$, (f) $\lambda=0.2324$.}
\label{eigmode_pco}
\end{figure}

We recall some results for simple, one-dimensional solitary pulses
with respect to the variation of $Pco$, as shown in Fig.
\ref{1D_bifur}.
The travelling speed of a {\it stable} solitary pulse decreases as
$Pco$ increases.
When $Pco$ is above some critical value ($4.599\times10^{-5}mbar$
in Fig. \ref{1D_bifur}), the medium can not support pulses any
more.
Checking the leading eigenvalues of the linearization of the
system equations we identify the bifurcation at which the pulse is
lost as a {\it turning point} (or saddle-node) instability: a
single eigenvalue crosses zero and becomes positive as the branch
of travelling pulses turns around and becomes unstable
(saddle-type).
In a two-dimensional ring-shaped domain, the shape and stability
of travelling pulses is affected by the presence of inert
boundaries, as shown in Fig. \ref{bifur_pco}.
The insets show the oxygen coverage profile corresponding to the
steady pulse at different locations along the diagram.
The inner and outer radius of the ring are fixed at dimensionless
values of 20 and 30 respectively (1 unit corresponds to
approximately 3.8 ${\mu}m$s).
Comparing the bifurcation diagrams in Fig. \ref{bifur_pco} and
\ref{1D_bifur} shows that the turning point bifurcation is still
present, but now we find that the pulse loses stability before the
turning point due to a Hopf bifurcation; simulations indicate that
this bifurcation is subcritical.
We plot the eigenmodes corresponding to the leading eigenvalues at
points $H$ and $p$ along the diagram in Fig. \ref{eigmode_pco}.
The leading eigenmodes at the Hopf bifurcation point $H$, Fig.
\ref{eigmode_pco}(b) and \ref{eigmode_pco}(c), constitute a
complex pair crossing the imaginary axis and rendering the pulse
unstable.
The spatial structure of these eigenmodes clearly resides close to
the inner boundary, suggesting that this boundary is responsible
for the instability.
An eigenvalue zero corresponding to the rotational invariance is
also found; its eigenvector (Fig. \ref{eigmode_pco}(a) and
\ref{eigmode_pco}(d)) can be obtained from the pulse shape by
differentiation with respect to $\theta$.
When the pulse propagates in the ring, its shape is no longer
planar (see the snapshot of a pulse on the stable branch in Fig.
\ref{bifur_pco}); it has to curve so as to satisfy the no-flux
boundary condition on both curved boundaries.
The end of the pulse close to the inner boundary takes a convex
shape and it is known that such a convex shape of pulses may lead
to instability when the curvature grows above some critical value
\cite{pulse2}.
We will see this boundary effect on the stability of pulses more
clearly below, when we present bifurcation diagrams from a direct
continuation in the ring radius.
After the Hopf bifurcation,  the eigenvalue pair responsible for
it collapses on the real axis, giving rise to two real positive
eigenvalues close to point $p$.
Fig. \ref{eigmode_pco}(e) and \ref{eigmode_pco}(f) show the
corresponding eigenmodes of these two real eigenvalues at point
$p$; the absolutely smaller one of the two proceeds to cross zero
at the turning point.

For the two-dimensional ring structure, the shape and stability of
the pulse solutions are dictated by the choice of $r_i$ and $r_o$.
For more general geometries (e.g. non-concentric boundaries) one
can parameterize the geometry through the inner curvature, the
outer curvature and the distance between the boundaries.
We will look at the effect of geometry taking two distinct
one-parameter paths: (a) change radii while keeping the ring width
fixed; and (b) fix the outer boundary and vary the inner one.

\subsection{Influence of varying the boundary curvature}

\begin{figure}
\centering
\includegraphics[width=0.9\columnwidth]{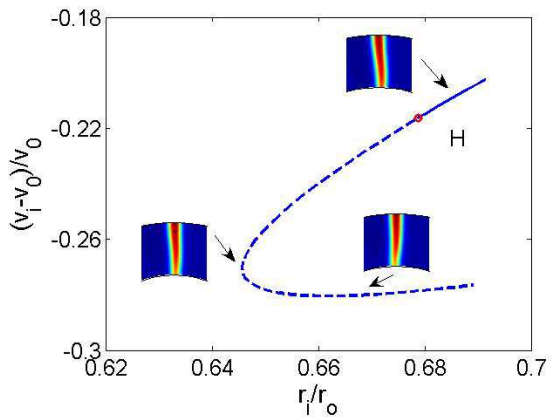}
\caption{Bifurcation diagram varying the boundary curvatures while
fixing ring width.
$v_i$ is the local velocity of the propagating pulse {\it at the
inner boundary}.
$v_0$ is the propagating speed of a one-dimensional pulse.
The increase of boundary curvature leads to a pulse instability
that is associated with a Hopf bifurcation, followed again by a
turning point.
$T= 540K, P_{O_2}= 1.33\times10^{-4} mbar,
P_{CO}=4.597\times10^{-5} mbar$. $r_o-r_i$ is fixed at 8.}
\label{bifur_w_8}
\end{figure}

\begin{figure}
\centering
\includegraphics[width=0.9\columnwidth]{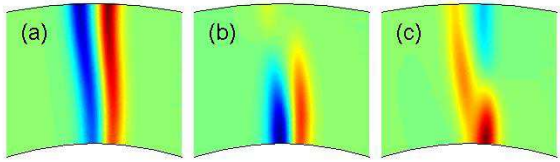}
\caption{Three leading eigenvalues and corresponding eigenmodes at
the Hopf bifurcation point $H$ in Fig. \ref{bifur_w_8}. (a)
$\lambda=-0.0012$, (b) and (c) $\lambda=-0.0025\pm0.2899i$.}
\label{eigmode_w}
\end{figure}

Figure \ref{bifur_w_8} reflects the influence of the curvature (of
both boundaries) on the pulse stability with $r_o-r_i$ held fixed.
%
%
The leading eigenvalues and corresponding eigenmodes at the Hopf
bifurcation point are plotted in Fig. \ref{eigmode_w}.
Similar to the first row of Fig. \ref{eigmode_pco}, the eigenmodes
corresponding to the eigenvalue pair crossing the imaginary axis
show some structure close to the inner boundary, but here both the
Hopf and SN bifurcation (at the turning point) are induced by the
change of boundary curvature alone.
The Hopf bifurcation is again subcritical (as simulations
indicate).

\begin{figure}
\centering
\includegraphics[width=0.9\columnwidth]{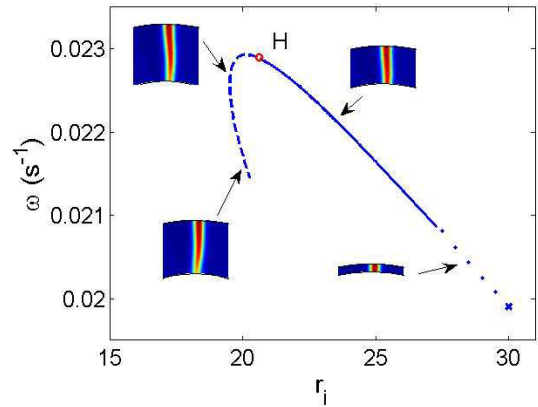}
\caption{Bifurcation diagram of pulses in a 2D ring showing the
influence of varying the inner ring radius $r_i$ on the travelling
pulse stability.
Point $H$, marked with a red circle, denotes a Hopf bifurcation.
Solid (dashed) lines denote stable (unstable) steady pulse
solutions.
For $r_i$ close to 30, steady pulse solutions are computed at
discrete points.
Insets show the profiles of O coverage.
The dimensionless outer radius of the ring $r_o$ is fixed at 30.
$T= 540K, P_{O_2}= 1.33\times10^{-4} mbar,
P_{CO}=4.597\times10^{-5} mbar$.} \label{bifur_ri}
\end{figure}

In the limit of $r_i/r_o=1$ as $r_i\rightarrow\infty$, we should
have an effectively one-dimensional planar pulse with a constant
velocity $v_0$.
As we decrease $r_i$, the local curvature of the pulse close to
the inner boundary increases and the local shape of the pulse
becomes more convex (compare the top inset in Fig. \ref{bifur_w_8}
with a planar pulse).
This is associated with a decreasing local velocity.
For a stable two-dimensional pulse expanding in an infinite medium
as a two-dimensional ring (with azimuthally uniform curvature),
the propagating velocity (and the pulse stability) increases as
the ring grows and the pulse becomes less convex \cite{pulse2}.
There also exists a critical curvature above which no pulse
solution exists.
A pulse ``fragment" propagating within a two-dimensional ring
structure acquires a non-uniform curvature, so as to satisfy the
no-flux boundary conditions applied on both curved boundaries (and
the linear speed distribution required for it to propagate
coherently).
This can lead to a locally concave shape of the pulse (larger
local velocity) close to the outer boundary and a locally convex
shape (smaller local velocity) close to the inner boundary.
The difference in the local curvature of the 2D pulse close to the
inert boundaries contributes to the spatial structure of the
unstable eigenmodes near the Hopf bifurcation point.
We now study how the pulse responds to changes in $r_i$ while
keeping $r_o$ fixed.
In this case, as we vary $r_i$, both the curvature of the inner
boundary and the distance between the two boundaries change.
A typical bifurcation diagram with respect to $r_i$ with fixed
$r_o$ is plotted in Fig. \ref{bifur_ri}.

When $r_i$ approaches the outer ring radius $r_o$, the pulse
becomes effectively one-dimensional and propagates at a limiting
velocity of approximately $0.6s^{-1}$ ($0.02s^{-1}$ for $\omega$
with $r_o=30$), consistent with the value read from the
one-dimensional pulse bifurcation diagram in Fig. \ref{1D_bifur}.
The qualitative structure of the bifurcation diagram in Fig.
\ref{bifur_ri} is similar to that in Fig. \ref{bifur_w_8}.

\begin{figure}
\centering
\includegraphics[width=0.9\columnwidth]{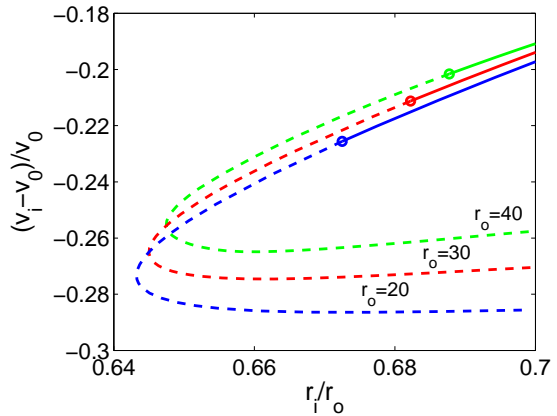}
\caption{Bifurcation diagram varying the inner ring radius $r_i$
while fixing $r_o$.
$v_i$ is the local velocity of the propagating pulse at the inner
boundary.
$v_0$ is the propagating speed of a one-dimensional pulse.
The $v_i$ associated with the Hopf bifurcation point depends on
$r_o$. } \label{vi_ri_ro}
\end{figure}

The continuation of pulse solutions with respect to $r_i$ for
several fixed values of $r_o$ is shown in Fig. \ref{vi_ri_ro}.
We see that, at criticality, the pulse speed at the inner boundary
(the slowest linear speed along the pulse) does depend on the
actual value of $r_o$ (and thus the overall pulse shape and its
nonuniform curvature).
%
%
All three stable branches for different $r_o$ converge to a
distinct point at $r_i/r_o=1$, that corresponds to the
one-dimensional problem (and so do, separately, all three unstable
ones).

When the diffusion coefficient is sufficient small, the medium can
no longer support a pulse solution; this is associated with a
turning point on the continuation curve with respect to the
diffusion coefficient at fixed $r_i$ and $r_o$ (not shown).
Scaling both ring radii by a factor of $k$ is equivalent to fixing
the geometry but solving a problem with a diffusion coefficient
scaled by a factor of $1/k^2$.
This can be used to rationalize the movement of the bifurcation
curves towards the right in Fig. \ref{vi_ri_ro}: Using
increasingly larger $r_o$ should lead to a critical $r_o$ above
which there exists no pulse solution for a given $r_i/r_o$.

\section{Dynamic simulations and experiments}
\label{experiments}

\begin{figure}
\centering
\includegraphics[width=0.9\columnwidth]{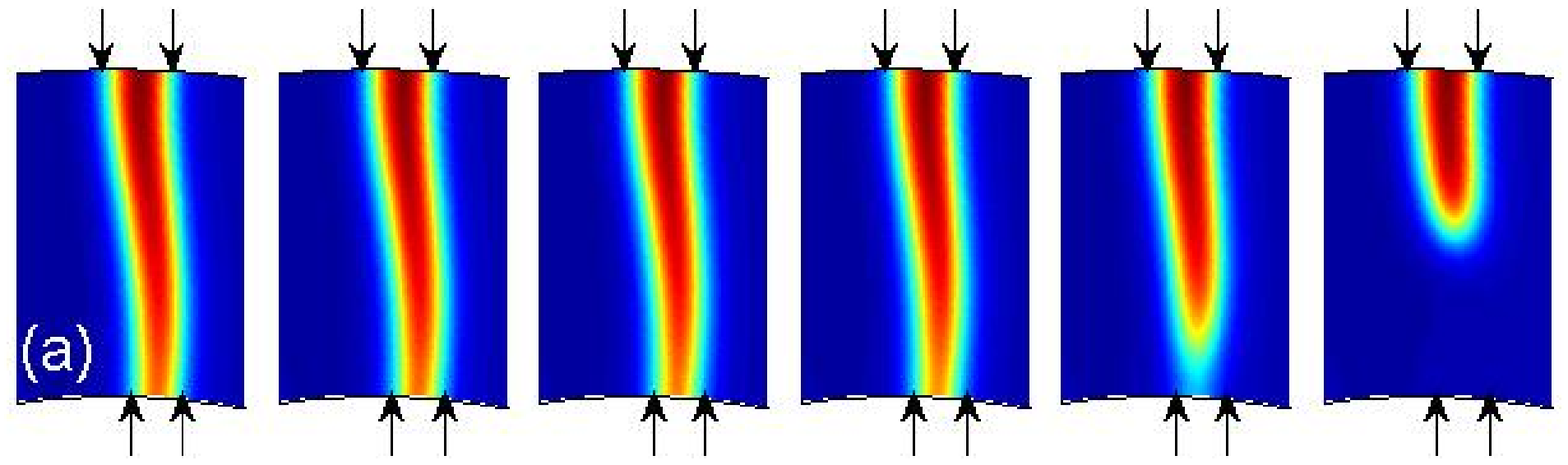}
\includegraphics[width=0.9\columnwidth]{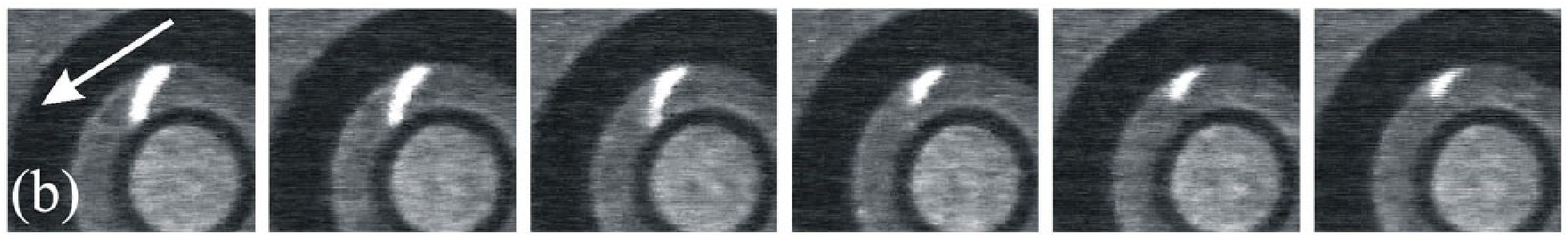}
\includegraphics[width=0.9\columnwidth]{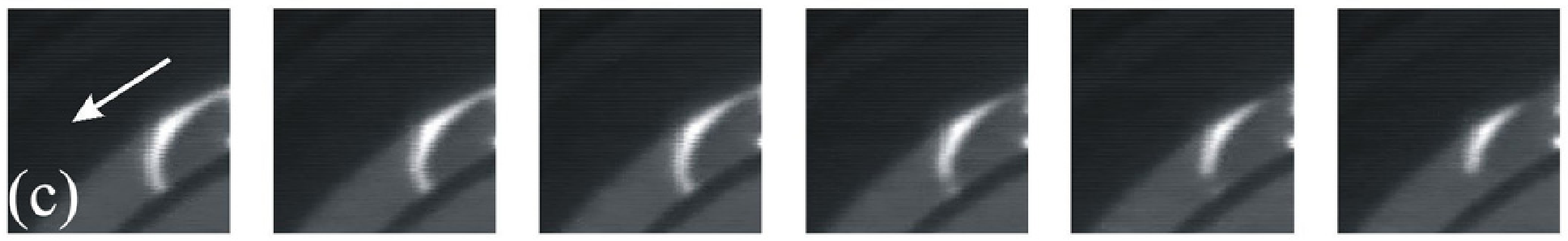}
\caption{Sequences of snapshots showing the decollation of
propagating pulses from the inner boundary of a ring structure.
(a) Numerical simulation. The time interval between each snapshot
is 2 seconds. Position-fixed arrows are used to show the change of
decollating pulse. The arrows are aligned to the boundary of the
pulse in the first snapshot. The ``tip" of the pulse at the inner
boundary first slowly moves toward the left and shrinkes at the
same time, then it decollates from the boundary. (b) and (c) are
experimental observations. The dark area is TiO$_2$ and the gray
area is Pt(110). The white propagating pulse is a CO-rich
one.Experimental conditions: $T= 447K, P_{O_2}= 4\times10^{-4}
mbar$,  (b) $P_{co}=3.35\times10^{-5} mbar$, (c)
$P_{co}=2.95\times10^{-5} mbar$.} \label{exp_decol}
\end{figure}

 We now consider how the boundary curvature-associated pulse instabilities manifest themselves in
time for both simulations and experiments.
A dynamic simulation close to (but just beyond) the Hopf
bifurcation point shows an initial ``tornado like" oscillation of
the ``tip" of the pulse close to inner boundary.
This tip becomes increasingly thinner (as indicated by the color
contours) and after some time the pulse ``decollates" from the
inner boundary and eventually dies (Fig. \ref{exp_decol}(a)).
This transient eventually evolves to the spatially uniform steady
state which is stable; thus the simulation is consistent with a
subcritical Hopf bifurcation.
The limit cycle branch is born unstable and ``backwards" in
parameter space; the saddle-type limit cycle and its stable
manifold constitute the separatrix between the stable pulse and
the stable uniform solution when these coexist.

Figure \ref{exp_decol}(b) and \ref{exp_decol}(c) show this
``decollation" mechanism from experimental observations of a
CO-rich pulse in a sequence of still shots 0.28 second apart.
The direction of motion of the pulse in the ring is denoted by an
arrow.
The data come from photoemission electron microscopy (PEEM)
observations of CO oxidation on micro-designed Pt catalysts; the
black areas are 2000 {\AA} tall Ti layers, deposited through
microlithography on the Pt(110) single crystal.
The inner and outer rings of the ring-shaped Pt ``corridor"
between the tall Ti ``walls" are 12.5 (160) and 20 (180) $\mu$m
respectively in Fig. \ref{exp_decol}(b) (Fig. \ref{exp_decol}(c)).
The experimental conditions are given in the figure caption; one
must remember, however, when observing these experiments that
diffusion on the Pt(110) crystal is {\it anisotropic}.
The ``thinning" and decollation of the pulse close to the inner
boundary can be clearly seen.

\section{Summary and Conclusions}
\label{conclusion}

We studied the effect of operating conditions and of geometry (in
particular, of boundary curvature) on the speed and stability of
pulses in excitable media; our illustrative model described CO
oxidation under high vacuum conditions on Pt(110).
The main instability is a decollation that occurs at the inner
ring boundary; while the turning point bounding one-dimensional
pulses still exists in full two-dimensional studies, the primary
destabilization occurs earlier, before the turning point itself.
We found that this primary instability involves a (subcritical)
Hopf bifurcation, leading to the decollation of the pulse from the
inner boundary, and (here) its eventual death.

In microdesigned/microstructured media, pulses often encounter
sudden changes in curvature (for example, at corners, or channel
exits); a ring geometry -more precisely, families of ring
geometries with varying radii- allows for the systematic study of
the effect of {\it steady}, controlled curvature.
This provides insight in the basic ``boundary decollation"
instability which underpins many of the events occurring in more
complex media.
Dynamic pattern formation in microchannels is studied in many
contexts beyond catalytic reactions (see e.g. recent work in
homogeneous reactions \cite{Kitahata}, and beyond reacting systems
in microfluidics \cite{Balazs2,Dreyfus}).
Boundary curvature and its variations is, we believe, a crucial
factor in all such applications.
Additional issues like boundary roughness (statistics of curvature
at a much finer scale) as well as the use of {\it active}
boundaries (e.g. consisting of a different catalyst, see
\cite{chaos}) will modify the results we described here.
We believe, however, that the basic reacting front decollation
from a curved boundary will be a recurring theme in all these
contexts, and an important component of pattern formation in
complex heterogeneous media.

{\bf Acknowledgements}. This work was partially supported by an
NSF/ITR grant and by AFOSR (IGK, LQ); LQ gratefully acknowledges
the support of a PPPL Fellowship.


\begin{thebibliography}{36}
\expandafter\ifx\csname
natexlab\endcsname\relax\def\natexlab#1{#1}\fi
\expandafter\ifx\csname bibnamefont\endcsname\relax
  \def\bibnamefont#1{#1}\fi
\expandafter\ifx\csname bibfnamefont\endcsname\relax
  \def\bibfnamefont#1{#1}\fi
\expandafter\ifx\csname citenamefont\endcsname\relax
  \def\citenamefont#1{#1}\fi
\expandafter\ifx\csname url\endcsname\relax
  \def\url#1{\texttt{#1}}\fi
\expandafter\ifx\csname
urlprefix\endcsname\relax\def\urlprefix{URL }\fi
\providecommand{\bibinfo}[2]{#2}
\providecommand{\eprint}[2][]{\url{#2}}

\bibitem[{\citenamefont{Cross and Hohenberg}(1993)}]{Cross}
\bibinfo{author}{\bibfnamefont{M.~C.} \bibnamefont{Cross}} \bibnamefont{and}
  \bibinfo{author}{\bibfnamefont{P.~C.} \bibnamefont{Hohenberg}},
  \bibinfo{journal}{Rev. Mod. Phys.} \textbf{\bibinfo{volume}{65}},
  \bibinfo{pages}{851} (\bibinfo{year}{1993}).

\bibitem[{\citenamefont{Imbihl}(2005)}]{Imbihl}
\bibinfo{author}{\bibfnamefont{R.}~\bibnamefont{Imbihl}},
  \bibinfo{journal}{Catalysis Today} \textbf{\bibinfo{volume}{105}},
  \bibinfo{pages}{206} (\bibinfo{year}{2005}).

\bibitem[{\citenamefont{Rotermund et~al.}(1990)\citenamefont{Rotermund, Engel,
  Kordesch, and Ertl}}]{peem}
\bibinfo{author}{\bibfnamefont{H.~H.} \bibnamefont{Rotermund}},
  \bibinfo{author}{\bibfnamefont{W.}~\bibnamefont{Engel}},
  \bibinfo{author}{\bibfnamefont{M.}~\bibnamefont{Kordesch}}, \bibnamefont{and}
  \bibinfo{author}{\bibfnamefont{G.}~\bibnamefont{Ertl}},
  \bibinfo{journal}{Nature (London)} \textbf{\bibinfo{volume}{343}},
  \bibinfo{pages}{355} (\bibinfo{year}{1990}).

\bibitem[{\citenamefont{Jiang et~al.}(2004)\citenamefont{Jiang, Bruzewicz,
  Wong, Piel, and Whitesides}}]{Whitesides}
\bibinfo{author}{\bibfnamefont{X.}~\bibnamefont{Jiang}},
  \bibinfo{author}{\bibfnamefont{D.~A.} \bibnamefont{Bruzewicz}},
  \bibinfo{author}{\bibfnamefont{A.~P.} \bibnamefont{Wong}},
  \bibinfo{author}{\bibfnamefont{M.}~\bibnamefont{Piel}}, \bibnamefont{and}
  \bibinfo{author}{\bibfnamefont{G.~M.} \bibnamefont{Whitesides}},
  \bibinfo{journal}{Proc. Nat. Acad. Sci. USA} \textbf{\bibinfo{volume}{102}},
  \bibinfo{pages}{975} (\bibinfo{year}{2004}).

\bibitem[{\citenamefont{Park et~al.}(1997)\citenamefont{Park, Harrison,
  Chaikin, Register, and Adamson}}]{Register}
\bibinfo{author}{\bibfnamefont{M.}~\bibnamefont{Park}},
  \bibinfo{author}{\bibfnamefont{C.}~\bibnamefont{Harrison}},
  \bibinfo{author}{\bibfnamefont{P.~M.} \bibnamefont{Chaikin}},
  \bibinfo{author}{\bibfnamefont{R.~A.} \bibnamefont{Register}},
  \bibnamefont{and} \bibinfo{author}{\bibfnamefont{D.~H.}
  \bibnamefont{Adamson}}, \bibinfo{journal}{Science}
  \textbf{\bibinfo{volume}{276}}, \bibinfo{pages}{1401} (\bibinfo{year}{1997}).

\bibitem[{\citenamefont{Verberg et~al.}(2004)\citenamefont{Verberg, Pooley,
  Yeomans, and Balazs}}]{Balazs1}
\bibinfo{author}{\bibfnamefont{R.}~\bibnamefont{Verberg}},
  \bibinfo{author}{\bibfnamefont{C.~M.} \bibnamefont{Pooley}},
  \bibinfo{author}{\bibfnamefont{J.~M.} \bibnamefont{Yeomans}},
  \bibnamefont{and} \bibinfo{author}{\bibfnamefont{A.~C.}
  \bibnamefont{Balazs}}, \bibinfo{journal}{Phys. Rev. Lett.}
  \textbf{\bibinfo{volume}{93}}, \bibinfo{pages}{184501}
  (\bibinfo{year}{2004}).

\bibitem[{\citenamefont{B\"oltau et~al.}(1998)\citenamefont{B\"oltau,
  Walheim, Mlynek, Krausch, and Steiner}}]{Boltau}
\bibinfo{author}{\bibfnamefont{M.}~\bibnamefont{B\"oltau}},
  \bibinfo{author}{\bibfnamefont{S.}~\bibnamefont{Walheim}},
  \bibinfo{author}{\bibfnamefont{J.}~\bibnamefont{Mlynek}},
  \bibinfo{author}{\bibfnamefont{G.}~\bibnamefont{Krausch}}, \bibnamefont{and}
  \bibinfo{author}{\bibfnamefont{U.}~\bibnamefont{Steiner}},
  \bibinfo{journal}{Nature (London)} \textbf{\bibinfo{volume}{391}},
  \bibinfo{pages}{877} (\bibinfo{year}{1998}).

\bibitem[{\citenamefont{Steinbock et~al.}(1995)\citenamefont{Steinbock,
  Kettunen, and Showalter}}]{Showalter}
\bibinfo{author}{\bibfnamefont{O.}~\bibnamefont{Steinbock}},
  \bibinfo{author}{\bibfnamefont{P.}~\bibnamefont{Kettunen}}, \bibnamefont{and}
  \bibinfo{author}{\bibfnamefont{K.}~\bibnamefont{Showalter}},
  \bibinfo{journal}{Science} \textbf{\bibinfo{volume}{269}},
  \bibinfo{pages}{1857} (\bibinfo{year}{1995}).

\bibitem[{\citenamefont{Graham et~al.}(1994)\citenamefont{Graham, Kevrekidis,
  Asakura, Lauterbach, Krischer, Rotermund, and Ertl}}]{boundaries2}
\bibinfo{author}{\bibfnamefont{M.~D.} \bibnamefont{Graham}},
  \bibinfo{author}{\bibfnamefont{I.~G.} \bibnamefont{Kevrekidis}},
  \bibinfo{author}{\bibfnamefont{K.}~\bibnamefont{Asakura}},
  \bibinfo{author}{\bibfnamefont{J.}~\bibnamefont{Lauterbach}},
  \bibinfo{author}{\bibfnamefont{K.}~\bibnamefont{Krischer}},
  \bibinfo{author}{\bibfnamefont{H.~H.} \bibnamefont{Rotermund}},
  \bibnamefont{and} \bibinfo{author}{\bibfnamefont{G.}~\bibnamefont{Ertl}},
  \bibinfo{journal}{Science} \textbf{\bibinfo{volume}{264}},
  \bibinfo{pages}{80} (\bibinfo{year}{1994}).

\bibitem[{\citenamefont{Tusscher and Panfilov}(2005)}]{Panfilov}
\bibinfo{author}{\bibfnamefont{K.~H. W. J.~T.} \bibnamefont{Tusscher}}
  \bibnamefont{and} \bibinfo{author}{\bibfnamefont{A.~V.}
  \bibnamefont{Panfilov}}, \bibinfo{journal}{Multiscale Model. Simul.}
  \textbf{\bibinfo{volume}{3}}, \bibinfo{pages}{265} (\bibinfo{year}{2005}).

\bibitem[{\citenamefont{Bub et~al.}(2005)\citenamefont{Bub, Shrier, and
  Glass}}]{Glass}
\bibinfo{author}{\bibfnamefont{G.}~\bibnamefont{Bub}},
  \bibinfo{author}{\bibfnamefont{A.}~\bibnamefont{Shrier}}, \bibnamefont{and}
  \bibinfo{author}{\bibfnamefont{L.}~\bibnamefont{Glass}},
  \bibinfo{journal}{Phys. Rev. Lett.} \textbf{\bibinfo{volume}{94}},
  \bibinfo{pages}{028105} (\bibinfo{year}{2005}).

\bibitem[{\citenamefont{Kaminaga et~al.}(2005)\citenamefont{Kaminaga, Vanag,
  and Epstein}}]{Epstein}
\bibinfo{author}{\bibfnamefont{A.}~\bibnamefont{Kaminaga}},
  \bibinfo{author}{\bibfnamefont{V.~K.} \bibnamefont{Vanag}}, \bibnamefont{and}
  \bibinfo{author}{\bibfnamefont{I.~R.} \bibnamefont{Epstein}},
  \bibinfo{journal}{Phys. Rev. Lett.} \textbf{\bibinfo{volume}{95}},
  \bibinfo{pages}{058302} (\bibinfo{year}{2005}).

\bibitem[{\citenamefont{Stone et~al.}(2004)\citenamefont{Stone, Stroock, and
  Ajdari}}]{Stone}
\bibinfo{author}{\bibfnamefont{H.~A.} \bibnamefont{Stone}},
  \bibinfo{author}{\bibfnamefont{A.~D.} \bibnamefont{Stroock}},
  \bibnamefont{and} \bibinfo{author}{\bibfnamefont{A.}~\bibnamefont{Ajdari}},
  \bibinfo{journal}{Annu. Rev. Fluid Mech.} \textbf{\bibinfo{volume}{36}},
  \bibinfo{pages}{381} (\bibinfo{year}{2004}).

\bibitem[{\citenamefont{Joanicot and Ajdari}(2005)}]{Joanicot}
\bibinfo{author}{\bibfnamefont{M.}~\bibnamefont{Joanicot}} \bibnamefont{and}
  \bibinfo{author}{\bibfnamefont{A.}~\bibnamefont{Ajdari}},
  \bibinfo{journal}{Science} \textbf{\bibinfo{volume}{309}},
  \bibinfo{pages}{887} (\bibinfo{year}{2005}).

\bibitem[{\citenamefont{Stroock et~al.}(2002)\citenamefont{Stroock, Dertinger,
  Ajdari, Mezi\'{c}, Stone, and Whitesides}}]{Stroock}
\bibinfo{author}{\bibfnamefont{A.~D.} \bibnamefont{Stroock}},
  \bibinfo{author}{\bibfnamefont{S.~K.~W.} \bibnamefont{Dertinger}},
  \bibinfo{author}{\bibfnamefont{A.}~\bibnamefont{Ajdari}},
  \bibinfo{author}{\bibfnamefont{I.}~\bibnamefont{Mezi\'{c}}},
  \bibinfo{author}{\bibfnamefont{H.~A.} \bibnamefont{Stone}}, \bibnamefont{and}
  \bibinfo{author}{\bibfnamefont{G.~M.} \bibnamefont{Whitesides}},
  \bibinfo{journal}{Science} \textbf{\bibinfo{volume}{295}},
  \bibinfo{pages}{647} (\bibinfo{year}{2002}).

\bibitem[{\citenamefont{Wolff et~al.}(2001)\citenamefont{Wolff, Papthanasiou,
  Kevrekidis, Rotermund, and Ertl}}]{Science2001}
\bibinfo{author}{\bibfnamefont{J.}~\bibnamefont{Wolff}},
  \bibinfo{author}{\bibfnamefont{A.~G.} \bibnamefont{Papthanasiou}},
  \bibinfo{author}{\bibfnamefont{I.~G.} \bibnamefont{Kevrekidis}},
  \bibinfo{author}{\bibfnamefont{H.~H.} \bibnamefont{Rotermund}},
  \bibnamefont{and} \bibinfo{author}{\bibfnamefont{G.}~\bibnamefont{Ertl}},
  \bibinfo{journal}{Science} \textbf{\bibinfo{volume}{294}},
  \bibinfo{pages}{134} (\bibinfo{year}{2001}).

\bibitem[{\citenamefont{Sakurai et~al.}(2002)\citenamefont{Sakurai, Mihaliuk,
  Chirila, and Showalter}}]{Sakurai}
\bibinfo{author}{\bibfnamefont{T.}~\bibnamefont{Sakurai}},
  \bibinfo{author}{\bibfnamefont{E.}~\bibnamefont{Mihaliuk}},
  \bibinfo{author}{\bibfnamefont{F.}~\bibnamefont{Chirila}}, \bibnamefont{and}
  \bibinfo{author}{\bibfnamefont{K.}~\bibnamefont{Showalter}},
  \bibinfo{journal}{Sience} \textbf{\bibinfo{volume}{296}},
  \bibinfo{pages}{2009} (\bibinfo{year}{2002}).

\bibitem[{\citenamefont{Lucchetta et~al.}(2005)\citenamefont{Lucchetta, Lee,
  Fu, Patel, and Ismagilov}}]{Lucchetta}
\bibinfo{author}{\bibfnamefont{E.~M.} \bibnamefont{Lucchetta}},
  \bibinfo{author}{\bibfnamefont{J.~H.} \bibnamefont{Lee}},
  \bibinfo{author}{\bibfnamefont{L.~A.} \bibnamefont{Fu}},
  \bibinfo{author}{\bibfnamefont{N.~H.} \bibnamefont{Patel}}, \bibnamefont{and}
  \bibinfo{author}{\bibfnamefont{R.~F.} \bibnamefont{Ismagilov}},
  \bibinfo{journal}{Nature} \textbf{\bibinfo{volume}{434}},
  \bibinfo{pages}{1134} (\bibinfo{year}{2005}).

\bibitem[{\citenamefont{Agladze et~al.}(1994)\citenamefont{Agladze, Keener,
  Muller, and Panfilov}}]{boundaries1}
\bibinfo{author}{\bibfnamefont{K.}~\bibnamefont{Agladze}},
  \bibinfo{author}{\bibfnamefont{J.~P.} \bibnamefont{Keener}},
  \bibinfo{author}{\bibfnamefont{S.~C.} \bibnamefont{Muller}},
  \bibnamefont{and} \bibinfo{author}{\bibfnamefont{A.}~\bibnamefont{Panfilov}},
  \bibinfo{journal}{Science} \textbf{\bibinfo{volume}{264}},
  \bibinfo{pages}{1746} (\bibinfo{year}{1994}).

\bibitem[{\citenamefont{Hartmann et~al.}(1996)\citenamefont{Hartmann,
  B\"ar, Kevrekidis, Krischer, and Imbihl}}]{boundaries3}
\bibinfo{author}{\bibfnamefont{N.}~\bibnamefont{Hartmann}},
  \bibinfo{author}{\bibfnamefont{M.}~\bibnamefont{B\"ar}},
  \bibinfo{author}{\bibfnamefont{I.~G.} \bibnamefont{Kevrekidis}},
  \bibinfo{author}{\bibfnamefont{K.}~\bibnamefont{Krischer}}, \bibnamefont{and}
  \bibinfo{author}{\bibfnamefont{R.}~\bibnamefont{Imbihl}},
  \bibinfo{journal}{Phys. Rev. Lett.} \textbf{\bibinfo{volume}{76}},
  \bibinfo{pages}{1384} (\bibinfo{year}{1996}).

\bibitem[{\citenamefont{B\"ar et~al.}(1996)\citenamefont{B\"ar, Bangia,
  Kevrekidis, Haas, Rotermund, and Ertl}}]{boundaries4}
\bibinfo{author}{\bibfnamefont{M.}~\bibnamefont{B\"ar}},
  \bibinfo{author}{\bibfnamefont{A.~K.} \bibnamefont{Bangia}},
  \bibinfo{author}{\bibfnamefont{I.~G.} \bibnamefont{Kevrekidis}},
  \bibinfo{author}{\bibfnamefont{G.}~\bibnamefont{Haas}},
  \bibinfo{author}{\bibfnamefont{H.~H.} \bibnamefont{Rotermund}},
  \bibnamefont{and} \bibinfo{author}{\bibfnamefont{G.}~\bibnamefont{Ertl}},
  \bibinfo{journal}{J. Phys. Chem.} \textbf{\bibinfo{volume}{100}},
  \bibinfo{pages}{19106} (\bibinfo{year}{1996}).

\bibitem[{\citenamefont{Tyson and Keener}(1988)}]{pulse1}
\bibinfo{author}{\bibfnamefont{J.~J.} \bibnamefont{Tyson}} \bibnamefont{and}
  \bibinfo{author}{\bibfnamefont{J.~P.} \bibnamefont{Keener}},
  \bibinfo{journal}{Physica D} \textbf{\bibinfo{volume}{32}},
  \bibinfo{pages}{327} (\bibinfo{year}{1988}).

\bibitem[{\citenamefont{Mikhailov}(1994)}]{pulse2}
\bibinfo{author}{\bibfnamefont{A.~S.} \bibnamefont{Mikhailov}},
  \emph{\bibinfo{title}{Foundations of Synergetics I: Distributed Active
  systems}} (\bibinfo{publisher}{Springer, New York}, \bibinfo{year}{1994}).

\bibitem[{\citenamefont{Krishnan et~al.}(1999)\citenamefont{Krishnan,
  Kevrekidis, Or-Guil, Zimmerman, and B\"ar}}]{pulse_1d1}
\bibinfo{author}{\bibfnamefont{J.}~\bibnamefont{Krishnan}},
  \bibinfo{author}{\bibfnamefont{I.~G.} \bibnamefont{Kevrekidis}},
  \bibinfo{author}{\bibfnamefont{M.}~\bibnamefont{Or-Guil}},
  \bibinfo{author}{\bibfnamefont{M.~G.} \bibnamefont{Zimmerman}},
  \bibnamefont{and}
  \bibinfo{author}{\bibfnamefont{M.}~\bibnamefont{B\"ar}},
  \bibinfo{journal}{Comput. Methods Appl. Mech. Engrg.}
  \textbf{\bibinfo{volume}{170}}, \bibinfo{pages}{253} (\bibinfo{year}{1999}).

\bibitem[{\citenamefont{Or-Guil et~al.}(2001)\citenamefont{Or-Guil, Krishnan,
  Kevrekidis, and B\"ar}}]{pulse_1d2}
\bibinfo{author}{\bibfnamefont{M.}~\bibnamefont{Or-Guil}},
  \bibinfo{author}{\bibfnamefont{J.}~\bibnamefont{Krishnan}},
  \bibinfo{author}{\bibfnamefont{I.~G.} \bibnamefont{Kevrekidis}},
  \bibnamefont{and}
  \bibinfo{author}{\bibfnamefont{M.}~\bibnamefont{B\"ar}},
  \bibinfo{journal}{Phys. Rev. E} \textbf{\bibinfo{volume}{64}},
  \bibinfo{pages}{046212} (\bibinfo{year}{2001}).

\bibitem[{\citenamefont{Krishnan et~al.}(2001)\citenamefont{Krishnan,
  Engelborghs, B\"ar, Lust, Roose, and Kevrekidis}}]{thin_ring}
\bibinfo{author}{\bibfnamefont{J.}~\bibnamefont{Krishnan}},
  \bibinfo{author}{\bibfnamefont{K.}~\bibnamefont{Engelborghs}},
  \bibinfo{author}{\bibfnamefont{M.}~\bibnamefont{B\"ar}},
  \bibinfo{author}{\bibfnamefont{K.}~\bibnamefont{Lust}},
  \bibinfo{author}{\bibfnamefont{D.}~\bibnamefont{Roose}}, \bibnamefont{and}
  \bibinfo{author}{\bibfnamefont{I.~G.} \bibnamefont{Kevrekidis}},
  \bibinfo{journal}{Physica D} \textbf{\bibinfo{volume}{154}},
  \bibinfo{pages}{85} (\bibinfo{year}{2001}).

\bibitem[{\citenamefont{B\"ar et~al.}(1994)\citenamefont{B\"ar,
  Gottschalk, Eiswirth, and Ertl}}]{two_variable_model}
\bibinfo{author}{\bibfnamefont{M.}~\bibnamefont{B\"ar}},
  \bibinfo{author}{\bibfnamefont{N.}~\bibnamefont{Gottschalk}},
  \bibinfo{author}{\bibfnamefont{M.}~\bibnamefont{Eiswirth}}, \bibnamefont{and}
  \bibinfo{author}{\bibfnamefont{G.}~\bibnamefont{Ertl}}, \bibinfo{journal}{J.
  Chem. Phys.} \textbf{\bibinfo{volume}{100}}, \bibinfo{pages}{1202}
  (\bibinfo{year}{1994}).

\bibitem[{\citenamefont{Krischer et~al.}(1992)\citenamefont{Krischer, Eiswirth,
  and Ertl}}]{full_model}
\bibinfo{author}{\bibfnamefont{K.}~\bibnamefont{Krischer}},
  \bibinfo{author}{\bibfnamefont{M.}~\bibnamefont{Eiswirth}}, \bibnamefont{and}
  \bibinfo{author}{\bibfnamefont{G.}~\bibnamefont{Ertl}}, \bibinfo{journal}{J.
  Chem. Phys.} \textbf{\bibinfo{volume}{96}}, \bibinfo{pages}{9161}
  (\bibinfo{year}{1992}).

\bibitem[{\citenamefont{Kelley}(1995)}]{Kelley}
\bibinfo{author}{\bibfnamefont{C.~T.} \bibnamefont{Kelley}},
  \emph{\bibinfo{title}{Iterative Methods for Linear and Nonlinear Equations}},
  vol.~\bibinfo{volume}{16} of \emph{\bibinfo{series}{Frontiers in Applied
  Mathematics}} (\bibinfo{publisher}{SIAM, Philadelphia},
  \bibinfo{year}{1995}).

\bibitem[{\citenamefont{Rowley and Marsden}(2000)}]{RowleyMarsden}
\bibinfo{author}{\bibfnamefont{C.~W.} \bibnamefont{Rowley}} \bibnamefont{and}
  \bibinfo{author}{\bibfnamefont{J.~E.} \bibnamefont{Marsden}},
  \bibinfo{journal}{Physica D} \textbf{\bibinfo{volume}{142}},
  \bibinfo{pages}{1} (\bibinfo{year}{2000}).

\bibitem[{\citenamefont{Rowley et~al.}(2003)\citenamefont{Rowley, Kevrekidis,
  Marsden, and Lust}}]{RKML}
\bibinfo{author}{\bibfnamefont{C.~W.} \bibnamefont{Rowley}},
  \bibinfo{author}{\bibfnamefont{I.~G.} \bibnamefont{Kevrekidis}},
  \bibinfo{author}{\bibfnamefont{J.~E.} \bibnamefont{Marsden}},
  \bibnamefont{and} \bibinfo{author}{\bibfnamefont{K.}~\bibnamefont{Lust}},
  \bibinfo{journal}{Nonlinearity} \textbf{\bibinfo{volume}{16}},
  \bibinfo{pages}{1257} (\bibinfo{year}{2003}).

\bibitem[{\citenamefont{Beyn and Th\"ummler}(2004)}]{Beyn}
\bibinfo{author}{\bibfnamefont{W.~J.} \bibnamefont{Beyn}} \bibnamefont{and}
  \bibinfo{author}{\bibfnamefont{V.}~\bibnamefont{Th\"ummler}},
  \bibinfo{journal}{SIAM J. Appl. Dyn. Syst.} \textbf{\bibinfo{volume}{3}},
  \bibinfo{pages}{85} (\bibinfo{year}{2004}).

\bibitem[{\citenamefont{Kitahata et~al.}(2005)\citenamefont{Kitahata, Yamada,
  Nakata, and Ichino}}]{Kitahata}
\bibinfo{author}{\bibfnamefont{H.}~\bibnamefont{Kitahata}},
  \bibinfo{author}{\bibfnamefont{A.}~\bibnamefont{Yamada}},
  \bibinfo{author}{\bibfnamefont{S.}~\bibnamefont{Nakata}}, \bibnamefont{and}
  \bibinfo{author}{\bibfnamefont{T.}~\bibnamefont{Ichino}},
  \bibinfo{journal}{J. Phys. Chem. A} \textbf{\bibinfo{volume}{109}},
  \bibinfo{pages}{4973} (\bibinfo{year}{2005}).

\bibitem[{\citenamefont{Kuksenok et~al.}(2003)\citenamefont{Kuksenok, Jasnow,
  Yeomans, and Balazs}}]{Balazs2}
\bibinfo{author}{\bibfnamefont{O.}~\bibnamefont{Kuksenok}},
  \bibinfo{author}{\bibfnamefont{D.}~\bibnamefont{Jasnow}},
  \bibinfo{author}{\bibfnamefont{J.}~\bibnamefont{Yeomans}}, \bibnamefont{and}
  \bibinfo{author}{\bibfnamefont{A.~C.} \bibnamefont{Balazs}},
  \bibinfo{journal}{Phys. Rev. Lett.} \textbf{\bibinfo{volume}{91}},
  \bibinfo{pages}{108303} (\bibinfo{year}{2003}).

\bibitem[{\citenamefont{Dreyfus et~al.}(2003)\citenamefont{Dreyfus, Tabeling,
  and Willaime}}]{Dreyfus}
\bibinfo{author}{\bibfnamefont{R.}~\bibnamefont{Dreyfus}},
  \bibinfo{author}{\bibfnamefont{P.}~\bibnamefont{Tabeling}}, \bibnamefont{and}
  \bibinfo{author}{\bibfnamefont{H.}~\bibnamefont{Willaime}},
  \bibinfo{journal}{Phys. Rev. Lett.} \textbf{\bibinfo{volume}{90}},
  \bibinfo{pages}{144505} (\bibinfo{year}{2003}).

\bibitem[{\citenamefont{Li et~al.}(2002)\citenamefont{Li, Kevrekidis, Pollmann,
  Papathanasiou, and Rotermund}}]{chaos}
\bibinfo{author}{\bibfnamefont{X.}~\bibnamefont{Li}},
  \bibinfo{author}{\bibfnamefont{I.~G.} \bibnamefont{Kevrekidis}},
  \bibinfo{author}{\bibfnamefont{M.}~\bibnamefont{Pollmann}},
  \bibinfo{author}{\bibfnamefont{A.~G.} \bibnamefont{Papathanasiou}},
  \bibnamefont{and} \bibinfo{author}{\bibfnamefont{H.~H.}
  \bibnamefont{Rotermund}}, \bibinfo{journal}{Chaos}
  \textbf{\bibinfo{volume}{12}}, \bibinfo{pages}{190} (\bibinfo{year}{2002}).

\end{thebibliography}

\end{document}